\documentclass[a4paper]{easychair}
\usepackage{fullpage}
\usepackage{microtype}
\usepackage[compact]{titlesec}

\usepackage{randtext}
\newcommand{\EmailAddress}[2]{\randomize{#1}@\randomize{#2}}

\definecolor{ltblue}{rgb}{0,0.4,0.4}
\definecolor{dkblue}{rgb}{0,0.1,0.6}
\definecolor{dkgreen}{rgb}{0,0.4,0}
\definecolor{dkviolet}{rgb}{0.3,0,0.5}
\definecolor{dkred}{rgb}{0.5,0,0}
\usepackage{listings}
\usepackage{lstcoq}

\usepackage[inline]{enumitem}
\setlist{nosep}

\title{Experience Report: Smuggling a Little Bit of Coq
  Inside a CAD Development Context (Extended Abstract)}
\author{Dimitur Nikolaev Krustev
\institute{
  IGE+XAO Balkan\\ Sofia, Bulgaria\\
  \email{\quad \EmailAddress{dkrustev}{ige-xao.com}}
}
}

\authorrunning{D.N.Krustev}
\titlerunning{Smuggling a Little Bit of Coq}

\begin{document}

\maketitle

\begin{abstract}
  While the use of formal verification techniques is well established in
  the development of mission-critical software, it is still rare in the 
  production of most other kinds of software.
  We share our experience that a formal verification tool such as Coq
  can be very useful and practical in the context of off-the-shelf software
  development -- CAD in particular -- at least in some occasions.
  The emphasis is on 3 main areas: 
  factors that can enable the use of Coq in an industrial context; 
  some typical examples of tasks, where Coq can offer an advantage; 
  examples of issues to overcome -- and some non-issues -- when integrating Coq 
  in a standard development process.
\end{abstract}

\section{Introduction}

A number of systems for formal verification  
-- Coq, Isabelle/HOL, HOL Light, Agda, Dafny, ACL2 \cite{PGL-045}, to name just a few --
can be used in software development.
In the context of Coq, many different methods exist for verifying software correctness, for example:
\begin{enumerate*}[label=\itshape\alph*\upshape)]
  \item built-in extraction mechanism: producing executable code from Coq definitions;
  \item producing Coq proof obligations from code in mainstream languages such as Java or C
    (using tools such as Why \cite{filliatre13esop} with different front-ends);
  \item using a library such as Bedrock \cite{10.1145/2544174.2500592} for verifying
    low-level and/or imperative code.
\end{enumerate*}
All these methods offer different trade-offs concerning learning curve, ease of use, power, etc.
Electrical CAD software encompasses a variety of tools -- schematic diagram editors,
calculation and simulation of electrical installations, part number/document/product life-cycle
management, etc.
They involve a combination of different technologies (2D/3D graphics, databases, web services),
and a mixture of complex algorithms (for automatic diagram layout, cable routing, etc.)
and domain-specific business rules.
In this context the most 
important consideration is not ensuring 100\% bug-free code, but optimizing 
the cost/quality ratio.
Besides, quality issues often stem from requirements, which are imprecise
and change over time.
In such a situation, any attempt at full-scale verification can fail just because
it is impossible or impractical to formally specify the final needs.
What we found useful was rather to detect small sub-tasks with 
the following combination of features:
\begin{enumerate*} [label=\itshape\alph*\upshape)]
  \item requirements are relatively easy to formally specify and stable over time;
  \item the task involves a complicated or non-standard algorithm, which, while
    small as code size, can have disproportionately high impact on the quality of the full product.
\end{enumerate*}
Even in such cases, we need to keep formal verification as lightweight as possible,
in order to justify it as a better alternative to other quality assurance techniques.
For now, we try to avoid the use of external tools or libraries not
coming with the standard Coq installation.
We rely on the built-in extraction to produce the executable code
we integrate in our products.
A major enabler for this approach is the growing popularity of functional programming
in industrial software development.
In our case, most of our new software development is based on .NET,
and we can integrate extracted OCaml code in F\# projects
(with minor tweaks).

\section{Tasks Suitable for Coq Verification}

So far, we have detected 3 main kinds of small tasks suitable to be implemented+verified in Coq:
\begin{itemize}
  \item Specialized or non-standard algorithms and data structures, without suitable
    existing implementations.
    Examples include some graph algorithms, such as length-preserving tree layout, A* search,
    branch-and-bound TSP; and specialized data structures, such as union-find, priority queues, etc.
  \item Some algorithms specific for the problem domain. 
    An example: how to apply a set of "patches" --
    each of them applicable for only a set of products -- to a set of data items, each of which is also 
    applicable only for a set of products. 
  \item Programming-language related tasks. 
    A bigger example consisted of a specialized in-house
    configuration language, which we had to extend with operations on collections, local definitions, and
    type inference.
    The type system we selected was based on Leijen's extensible records \cite{Leijen2005ExtensibleRW}.
    What we did verify is the semantics preservation of a translation from our language to
    a variant of Leijen's core language.
    We used Ott \cite{10.1145/1291220.1291155} with Coq, in order 
    to produce nice-looking documentation.
\end{itemize}
While the last case is a well-established application area, it was rather an exception in our setting.
In our estimation, specialized algorithms and data structures with stable simple
specifications, as described in the first case, are where the use of 
Coq is most beneficial in an industrial setting.

\section{Case Study: A* Search}

Our in-house implementation of A* search \cite{RussellNorvigAIMA} demonstrates the benefits of 
limited formal verification.
While well known, this algorithm is not typically found in standard libraries.
Besides, we needed our own implementation for a couple of reasons:
its intended uses required a very generic API (arbitrary types for route weight, possibility
to have multiple targets, etc.);
we wanted to be able to fine-tune the performance, for example by using LIFO when resolving ties.
While very small as code size (Coq specifications - 173 lines, proofs - 203, comments - 29; extracted F\# - 39), 
the algorithm relies for its correctness on some non-trivial invariants.
We estimate that ensuring the correctness of such code by more traditional means (different forms of testing),
would result in a much greater total cost than what we spent in verification.
The code is used in a much larger module (108 KLOC), part of an even larger final product,
which underwent extensive testing on all levels for a couple of years, with more than 700 issues
detected (both implementation problems and requirement changes).
None of these concerned our A* implementation, even though it is executed
multiple times during each product test,
thus proving our approach to be not only economical, but also effective.
We have only verified that our implementation of A* search produces a valid path,
without checking that this path is optimal.
This is another example of a practical trade-off: only spend verification time on what is most
important -- in the given context -- for the quality of the final product.

F\# and OCaml share a common core, but there are key differences when going beyond 
the core, such as the less expressive F\# module system.
The lack of higher-kinded types in F\# can also make code extracted from Coq not directly usable.
Those limitations are not critical for us most of the time:
we usually only verify small fragments of code, and they change very infrequently, if at all.
As a result, we can afford to: 
a) manually copy the extracted code to the corresponding part of
F\# sources, without spending time on integrating Coq in our build process; 
b) perform some minimal manual modifications to make the extracted code compile, if necessary.
Still, some small tricks can make the extracted F\# code easier to use.
For example, in the A* implementation, we need to use finite sets with a particular type of elements.
We prefer to rely on the finite-set implementation of the F\# standard library, so
we have made a small abstraction to make this work:
\begin{coqka}
Record FinSetOps (A: Set) := { FinSet: Set; empty: FinSet;
  add: forall (A_dec: forall x y: A, {x = y} + {x <> y}), A -> FinSet -> FinSet;
  contains: forall (A_dec: forall x y: A, {x = y} + {x <> y}), A -> FinSet -> bool;  ... }.
Variable fsOps: FinSetOps Node.
\end{coqka}
This results in extracted code such as \verb|let closedSet' = fsOps.add node_dec node closedSet|,
which we can easily modify to a form, which can compile: \verb|let closedSet' = Set.add node closedSet|.

\paragraph{Summary}
The A* search implementation, as well as several similar examples, have convinced us
that the verification -- in Coq -- of even tiny fragments of big off-the-shelf software
can provide substantial benefits.
We plan to continue expanding such experiments in the future.

\label{sect:bib}
\bibliographystyle{plain}
\bibliography{CoqExperience}

\end{document}